\documentclass[iop]{emulateapj}

\usepackage{natbib}

\usepackage{color}

\newcommand{\eg}{e.g., }
\newcommand{\ie}{i.e., }
\newcommand{\Msun}{M_{\odot}}
\newcommand{\Md}{M_{\rm dust}}
\newcommand{\Td}{T_{\rm dust}}
\newcommand{\kms}{km~s$^{-1}$}

\newcommand{\mum}{$\mu$m}
\newcommand{\Mdot}{\dot{M}}
\def\gsim{\mathrel{\rlap{\lower 4pt \hbox{\hskip 1pt $\sim$}}\raise 1pt
\hbox {$>$}}}
\def\lsim{\mathrel{\rlap{\lower 4pt \hbox{\hskip 1pt $\sim$}}\raise 1pt
\hbox {$<$}}}

\newcommand{\vsn}{v_{\rm SN}}
\newcommand{\vwind}{v_{\rm wind}}

\def\ion#1#2{{\rm #1}~{\sc #2}}

\newcommand{\nar}{NewA Rev.}

\shorttitle{IR Emission from Core-Collapse Supernovae at the Transitional Phase}
\shortauthors{Tanaka et al.}

\begin{document}

\title{A Search for Infrared Emission from Core-Collapse Supernovae at the Transitional Phase}
\author{
Masaomi Tanaka\altaffilmark{1,2}, 
Takaya Nozawa\altaffilmark{2}, 
Itsuki Sakon\altaffilmark{3}, 
Takashi Onaka\altaffilmark{3}, 
Ko Arimatsu\altaffilmark{3}, 
Ryo Ohsawa\altaffilmark{3}, 
Keiichi Maeda\altaffilmark{2}, 
Takehiko Wada\altaffilmark{4}, 
Hideo Matsuhara\altaffilmark{4}, and
Hidehiro Kaneda\altaffilmark{5}
}

\altaffiltext{1}{National Astronomical Observatory of Japan, Mitaka, Tokyo, Japan; masaomi.tanaka@nao.ac.jp} 
\altaffiltext{2}{Institute for the Physics and Mathematics of the Universe, University of Tokyo, Kashiwa, Japan; 
takaya.nozawa@ipmu.jp}
\altaffiltext{3}{Department of Astronomy, Graduate School of Science, University of Tokyo, Bunkyo-ku, Tokyo, Japan}
\altaffiltext{4}{Institute of Space and Astronautical Science, Japan Aerospace Exploration Agency, Sagamihara, Kanagawa, Japan}
\altaffiltext{5}{Graduate School of Science, Nagoya University, Chikusa-ku, Nagoya, Japan}

\begin{abstract}
Most of the observational studies of supernova (SN) explosions 
are limited to early phases 
($<$ a few yr after the explosion) of extragalactic SNe
and observations of SN remnants ($>$ 100 yr)
in our Galaxy or very nearby galaxies.
SNe at the epoch between these two, which we call ``transitional'' phase,
have not been explored in detail except for several extragalactic SNe 
including SN 1987A in the Large Magellanic Cloud.
We present theoretical predictions for the infrared (IR) dust emissions by
several mechanisms; emission from dust formed in the SN ejecta,
light echo by circumstellar and interstellar dust, and 
emission from shocked circumstellar dust.
We search for IR emission from 6 core-collapse SNe at the transitional phase
in the nearby galaxies NGC 1313, NGC 6946, and M101
by using the data taken with the AKARI satellite 
and {\it Spitzer}.
Among 6 targets, we detect the emission from SN 1978K in NGC 1313.
SN 1978K is associated with $1.3 \times 10^{-3} \Msun$ of silicate dust.
We show that, among several mechanisms, 
the shocked circumstellar dust is the most probable emission source to
explain the IR emission observed for CSM-rich SN 1978K.
IR emission from the other 5 objects is not detected.
Our current observations are sensitive to IR luminosity of 
$> 10^{38} {\rm erg\ s^{-1}}$,
and the non-detection of SN 1962M
excludes the existence of the shocked circumstellar 
dust for a high gas mass-loss rate of $\sim 10^{-4}\ \Msun\ {\rm yr^{-1}}$.
Observations of SNe at the transitional phase with future IR satellites
will fill the gap of IR observations 
of SNe with the age of $10-100$ years, and 
give a new opportunity to study the circumstellar and 
interstellar environments of the progenitor, and possibly 
dust formation in SNe.
\end{abstract}


\section{Introduction}
\label{sec:intro}

Core-collapse supernovae (SNe) are the explosions of 
massive stars at the end of their lives.
SN ejecta expand into the interstellar medium (ISM)
or circumstellar medium (CSM) with a huge kinetic energy ($\sim 10^{51}$ erg).
First, SN ejecta experience a free expansion.
As the ejecta sweep up the surrounding material, 
the SN begins to decelerate.
When the swept-up mass becomes comparable to the mass of the SN ejecta
at $t \sim$ a few hundreds yr 
(hereafter $t$ denotes the time after the explosion),
the SN ejecta expand self-similarly (so-called Sedov phase).
Finally, the SN merges into the ISM at $t \gsim 10,000$ yr.

Observational studies of SNe
have been performed mainly by the two distinct ways.
One is the observations of extragalactic SNe
at their early phases ($t < 10$ yr),
and the other is observations of supernova remnants 
(SNRs, $t \gsim$ 100 yr) in our Galaxy or very nearby galaxies.
In contrast, SNe at the epoch between these two ($t=10-100$ yr), 
which we call ``transitional'' phase, have not been explored in detail.
At such epochs, only a few objects have been studied. 
The best case is SN 1987A in the Large Magellanic Cloud 
\citep[and references therein]{mccray07}.
Other examples include several X-ray SNe 
(\eg \citealt{immler05,immler0579C,soria08}), 
and long-lasting SNe, such as 
SNe 1978K \citep{ryder93,schlegel99,schlegel00,schlegel04,lenz07,smith07}
and 1988Z \citep{williams02,schlegel06}.
See also a recent paper on SN 1980K by \citet{sugerman12}.

This transitional phase is important to understand 
a long term evolution of SNe, from SN to SNR.
A recent study by \citet{larsson11} shows that 
SN 1987A, in fact, experiences a transition from SN to SNR in this phase.
However, since most observations of SNe at transitional phase
except for SN 1987A have been performed at X-ray or 
radio wavelengths, the entire properties of 
SNe at the transitional phase are not fully understood.

\begin{deluxetable*}{cclcllc} 
\tablewidth{0pt}
\tablecaption{List of Targets}
\tablehead{
Host Galaxy & 
Distance & 
\multicolumn{1}{c}{SN}  & 
SN Type & 
\multicolumn{2}{c}{Position} & 
Position Ref. \\
   & 
(Mpc) &
 &
 &
\multicolumn{1}{c}{$\alpha$ (J2000.0)} &
\multicolumn{1}{c}{$\delta$ (J2000.0)} &
}
\startdata
NGC 1313        & $4.13 \pm 0.11$ 
                & SN 1962M  & II  & 03h18m12.2s & $-$66$^{\circ}$31$'$38$''$   & 1  \\
             &  & SN 1978K  & IIn & 03h17m39.0s & $-$66$^{\circ}$33$'$048$''$     & 2 \\
NGC 6946        & $5.6 \pm 0.3$ 
                & SN 1917A  & II  & 20h34m46.90s & +60$^{\circ}$07$'$29.08$''$    & 3  \\
             &  & SN 1968D  & II  & 20h34m58.41s & +60$^{\circ}$09$'$34.48$''$   & 4  \\
M101            & $6.7 \pm 0.3$
                & SN 1909A\tablenotemark{a} & II peculiar & 14h02m03.1s  & +54$^{\circ}$27$'$58$''$ &  3  \\
             &  & SN 1951H  & II  & 14h03m55.3s & +54$^{\circ}$21$'$41$''$ & 3   \\
\enddata
\tablenotetext{a}{SN 1909A is out of field of view in the IRC images, 
and only in the Spitzer/MIPS image.}
\tablerefs{(1) \citealt{artyukhina96}, (2)\citealt{dopita90}, 
(3) \citealt{barbon99}, (4) \citealt{vandyk94}}
\label{tab:targets}
\end{deluxetable*}

\begin{deluxetable*}{ccccccc} 
\tablewidth{0pt}
\tablecaption{Log of AKARI/IRC Observations}
\tablehead{
Galaxy &  
\multicolumn{2}{c}{Date} &  
\multicolumn{2}{c}{Observation ID} &
\multicolumn{2}{c}{Exposure Time (s)} 
\\
  & 
NIR, MIR-S & 
MIR-L & 
NIR, MIR-S & 
MIR-L &
NIR &
MIR-S, L 
}
\startdata
NGC 1313        & 2006 Dec 3   & 2006 Dec 4   & 1400416-001  & 1400417-001 &  178  & 196 \\
NGC 6946        & 2006 Dec 18  & 2007 Jun 16  &  1400620-001 & 1402217-001 &  178  & 196   \\
M101            & 2007 Jun 14  & 2007 Jun 17  & 1402211-001  & 1402212-001 &  178  & 196  \\
\enddata
\label{tab:obslog}
\end{deluxetable*}

In this paper, we present the results of our search for 
infrared (IR) emission from core-collapse SNe at the transitional phase.
IR emission from SNe is expected to arise from 
dust associated with SNe.
There are 3 possible populations of dust grains, 
according to their locations;
(1) dust formed in the ejecta of SNe (SN dust), 
(2) dust formed by the pre-SN stellar 
wind, which is now located in the circumstellar region (CS dust), 
and (3) interstellar dust (IS dust).
CS and IS dust can contribute to the IR emission by absorbing 
the radiation from SNe and reemitting it (light echo).
CS dust is also heated and can also emit IR emission 
when it is swept up by the SN shock.
In these ways, dust grains in and around SNe convert the large kinetic energy 
and luminosity of the SNe into the IR luminosity.
Thus, IR observations of SNe can be used as a probe of the 
progenitor environment.

In Section \ref{sec:obs}, we describe the target selection,
the data used in the analysis, and results of photometry.
We use the data taken with two IR satellites,
AKARI satellite \citep{murakami07} and {\it Spitzer} \citep{werner04}, 
which provide the deepest images in near-IR (NIR) and mid-IR (MIR) wavelengths.
Dust models used in this paper are shown in Section \ref{sec:model}.
Among 6 targets, we detect IR emission from SN 1978K.
The origin of this emission is discussed in Section \ref{sec:1978K}.
In Section \ref{sec:normalSN}, 
theoretical predictions for the IR emission 
from normal SNe at the transitional phase are summarized, and 
implications of non-detection of 5 SNe are discussed.
Prospects for future observations are discussed in 
Section \ref{sec:future}.
Finally, we summarize the conclusions in Section \ref{sec:conclusions}.


\section{Targets and Observations}
\label{sec:obs}

\begin{deluxetable*}{cccccc} 
\tablewidth{0pt}
\tablecaption{Summary of Spitzer/MIPS Data}
\tablehead{
Galaxy & 
Date &
Exposure Time (s) &
Observing mode &
AOR key &
Program ID 
}
\startdata
NGC 1313                   & 2007 Sep 29  & 42  & Medium Scan & 22618112 & 40204: Kennicut et al. \\
NGC 6946                   & 2007 Jul 10  & 493 & Photometry  & 18271232 & 30494: Sugerman et al. \\
M101                       & 2008 Jan 5   & 42 & Medium Scan & 21380144 & 40352: Rieke et al.  \\\
\enddata
\label{tab:MIPS}
\end{deluxetable*}

\begin{deluxetable*}{lcccccccc} 
\tablewidth{0pt}
\tablecaption{Summary of Photometry}
\tablehead{
\multicolumn{1}{c}{SN} & 
Age \tablenotemark{a}& 
\multicolumn{7}{c}{Flux (mJy)\tablenotemark{b}} \\
   & 
(yr)&
IRC/N3 &
IRC/N4 &
IRC/S7 &
IRC/S11 &
IRC/L15 &
IRC/L24 &
MIPS/24\mum
}
\startdata
SN 1962M &  44.0   & $<$0.2        &  $<$0.2       &  $<$0.2        &  $<$0.2      &  $<$0.9      & $<$2        & ...  \\
         &  44.9   &  ...          &  ...          &  ...           &  ...         &  ...         & ...         & $<$1  \\
SN 1978K &  28.4   & 0.074 (0.014) & 0.086 (0.010) & 0.287 (0.039)  &  2.85 (0.24) & 2.53 (0.13)  & 2.73 (0.20) & ...   \\
         &  29.2   & ...           &  ...          &  ...           &  ...         & ...          & ...         & 3.06 (0.40) \\
SN 1917A &  89.5   & $<$0.1        & $<$0.2        &  $<$8          &  $<$6        & ...          & ...         & ... \\
         &  90.0   & ...           & ...           & ...            & ...          & $<$10        & $<$10       & ... \\
         &  90.0   & ...           & ...           & ...            & ...          & ...          & ...         & $<$10 \\
SN 1968D &  38.8   & $<$0.3        & $<$0.3        & $<$4           &  $<$5        & ...          & ...         & ... \\
         &  39.3   & ...           & ...           & ...            & ...          & $<$6         & $<$8        & ...\\
         &  39.4   & ...           & ...           & ...            & ...          & ...          & ...         & $<$5 \\
SN 1909A &  99.0   & ...           & ...           & ...            & ...          & ...          & ...         & $<$0.7 \\
SN 1951H &  55.9   & $<$0.07       &  $<$0.1       & $<$0.4         &  $<$0.6      &  $<$1        & $<$3        & ... \\       
         &  56.3   & ...           & ...           & ...            & ...          & ...          & ...         & $<$1 \\
\enddata
\tablenotetext{a}{
Age is given as year after the discovery, which
is a sound approximation for the age after the explosion.}
\tablenotetext{b}{
Values in parenthesis for SN 1978K represents $1\sigma$ error 
including the measurement error and the calibration error (5 $\%$).
In the case of non-detection, $3\sigma$ upper limit is given.}
\label{tab:phot}
\end{deluxetable*}

\subsection{Targets}

To search for IR emission from SNe at the transitional phase,
we target 7 core-collapse SNe, namely
SNe 1909A, 1917A, 1951H, 1962M, 1968D, and 1978K,
discovered in three very nearby galaxies NGC 1313, NGC 6946, and M101.
These targets are selected based on 
the proximity to the host galaxy and 
the availability of NIR and MIR imaging data 
taken with the IRC \citep{onaka07} 
onboard the AKARI satellite \citep{murakami07}.
General information on our targets is summarized in 
Table \ref{tab:targets}.

Note that SN 1948B (Type II) in NGC 6946 is excluded 
because of the uncertainty in its position.
SNe 1939C and 1969P in NGC 6946 
are also excluded since SN 1939C is classified as merely Type I 
(either core-collapse SNe or Type Ia SNe)
and the type of SN 1969P is unknown.
SN 1980K is also located in NGC 6946.
But since it is found that AKARI data are not as deep as 
the data presented by \citet{sugerman12},
we do not include SN 1980K.
SN 1970G (Type IIL) in M101 is excluded because 
the SN position is heavily contaminated by the \ion{H}{II}
region NGC 5455 \citep{fesen93}.

For the distance to NGC 1313, we adopt $4.13 \pm 0.11$ Mpc,
which is estimated from the tip of the red giant branch
\citep{mendez02}.
We use this value since it was also used by \citet{lenz07} 
and \citet{smith07} for the study of SN 1978K.
For NGC 6946, there are several distance measurements.
We adopt the average value of $5.6 \pm 0.3$ Mpc
from the the \ion{H}{I} Tully-Fisher relation \citep{pierce94},
the CO Tully-Fisher relation \citep{schoniger94},
and the expanding photosphere method of Type II SN 1980K
\citep{schmidt94} and SN 2004et \citep{sahu06}.
The distance to M101 is assumed to be $6.7 \pm 0.3$ Mpc,
which is estimated by the Cepheid calibration \citep{freedman01}.

\subsection{AKARI IRC Data}

The AKARI/IRC equips three channels, \ie 
NIR (1.8$-$5.5 \mum), MIR-S (4.6$-$13.4 \mum) 
and MIR-L (12.6$-$26.5 \mum).
Each channel has about a $10'$ $\times$ $10'$ field of view.
The NIR and MIR-S share the same field of view 
while the MIR-L observes a sky about 25' away.
Thus, the data we use in this paper consist of
two observational sequences.
The pixel scales of the detectors are 
$1.46''$ $\times$ $1.46''$, $2.34''$ $\times$ $2.34''$,
and $2.51''$ $\times$ $2.39''$ for the NIR, MIR-S and MIR-L,
respectively.

We use the archived imaging data taken in the AKARI 
mission program "ISM in our Galaxy and Nearby galaxies"
(ISMGN; \citealt{kaneda09}).
The data are taken with the two-filter mode 
[Astronomical Observation Template (AOT) IRC02,
see \citet{onaka07}].
We use 6 band images in total, \ie
N3 (reference wavelength 3.2 \mum), N4 (4.1 \mum),
S7 (7.0 \mum), S11 (11.0 \mum), L15 (15.0 \mum),
and L24 (24.0 \mum).
The wavelength coverage with the response larger than
$1/e$ of the peak is 2.7$-$3.8 \mum, 3.6$-$5.3 \mum, 
5.9$-$8.4 \mum, 8.5$-$13.1 \mum, 12.6$-$19.4 \mum,
and 20.3$-$26.5 \mum\ for N3, N4, S7, S11, L15, and L24,
respectively.
For more details of the IRC instrument, see \citet{onaka07}.
The log of observations is summarized in Table \ref{tab:obslog}.

The data were reduced by using the IRC imaging pipeline 
software version 20091022 \citep[see][for details]{lorente07}.
The pipeline performs the basic reduction, \eg
correction of bad pixels, subtraction of dark current, 
rejection of cosmic rays, 
linearity correction, flat fielding, 
and co-adding of the frames.

To determine the accurate position,
we match the sources in the N3, N4 and S7 band images
with those in Two Micron All Sky Survey catalog \citep{skrutskie06}.
The uncertainty in the position is $<2''$ for
the N3 and N4 band images, and  $<3''$ for the S7 band images.
Then, the S11, L15 and L24 images are matched 
with S7 images.
In these images, the uncertainty in position is estimated as $\sim 3''$, 
which is larger than that in the N3, N4 and S7 images 
because the matching is based on a small number of common sources
in the field of view.

To obtain the flux densities of the sources,
we perform aperture photometry
with the aperture of 10 pixel and 7.5 pixel for
the NIR and MIR-S/L bands, respectively, 
As the background, we use the annulus of 5 pixel width
just outside of the aperture radius.
When the SN is located at the edge of the image,
the aperture correction is applied.
The flux is calibrated by following \citet{lorente07}.
Color correction is {\it not} performed to the derived 
photometry because the intrinsic spectral shape 
of the source is not known.
Instead, we apply correction to the model when it is compared with
the observations (see Section \ref{sec:model}).

When the SN is not detected, we derive an 
upper limit of the flux by putting an artificial source
and measuring the flux of the source.
For this purpose, we need a point spread function (PSF) of each image.
For the N3 and N4 images, PSF is constructed 
by the point sources in the same image.
Since there are not many sources in the S7, S11, L15 and 
L24 images, we use an average PSF constructed
from sources in different fields \citep{arimatsu11}.

\begin{figure*}
\begin{center}
\includegraphics[angle=270,scale=0.85]{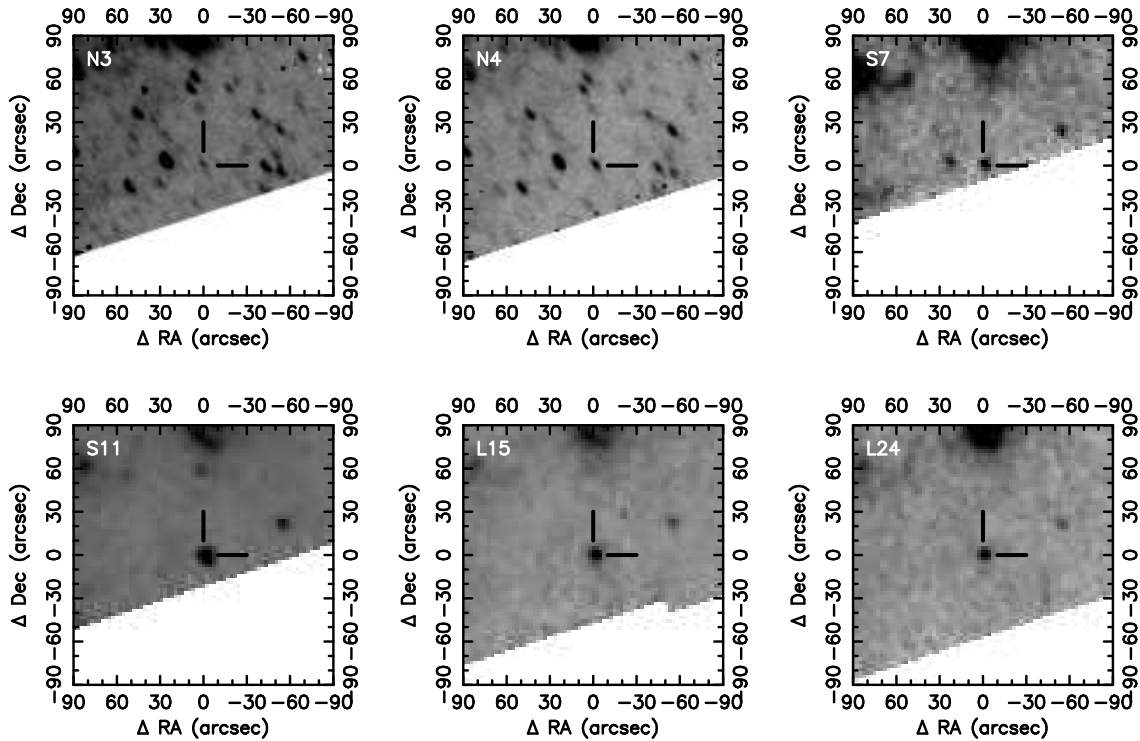}
\caption{
AKARI/IRC images around SN 1978K
in NGC 1313 (180'' $\times$ 180'' section, 
which corresponds to about 3.6 kpc $\times$ 3.6 kpc).
North is up, and east is to the left.
Note that the PSF of the N3 and N4 bands are 
slightly elongated in images on orbit due to a misalignment of 
the telescope and color aberrations in the NIR channel \citep{kaneda07}.
\label{fig:1978K}}
\end{center}
\end{figure*}

\subsection{Spitzer MIPS Data}

In addition to the AKARI/IRC data, we use archive data
taken with the Multiband Imaging Photometer for Spitzer \citep[MIPS,][]{rieke04}
onboard {\it Spitzer} \citep{werner04}.

All the 6 SNe are within the field of view 
of some observations with the MIPS 24 \mum\ band.
NGC 1313 was observed with the scan mode in 
the Spitzer Local Volume Legacy program \citep{dale09}.
M101 was also observed with the scan mode
by \citet{gordon08}.
SNe 1917A and 1968D in NGC 6946 are within 
the field of SN 2002hh \citep{meikle06}.
A summary of the data, including 
the Astronomical Observation Request (AOR) keys,
Program IDs and PI names, is shown in Table \ref{tab:MIPS}.

We use the post-basic calibrated data (PBCD) products,
which are produced by the Spitzer pipeline.
The pixel scale of the final image is $2.45''$ $\times$ $2.45''$.
The reference wavelength of the MIPS  24 \mum\ band is
23.68 \mum\ and the wavelength coverage is 
20.5 \mum\ -- 28.5 \mum\ 
(the range where the response is $>10\%$ of the peak).
The positional uncertainty is $1.4''$.

For photometry, we perform aperture photometry.
When the SN is not detected, we derive an upper limit 
of the flux by putting an artificial source 
around the SN position.
We constructed a PSF from point sources in the scan mode images
of NGC 1313 and M101 and use it to derive the upper limits.

The Spitzer/MIPS data are found to be 
slightly deeper than the AKARI/IRC data at 24 \mum\ 
(Section \ref{sec:results}).
Thus, we use MIPS data for the flux of 24 \mum.

\subsection{Results of Photometry}
\label{sec:results}

We detect IR emission associated with SN 1978K 
in all the bands (Table \ref{tab:phot}).
Figure \ref{fig:1978K} shows the AKARI/IRC images
around SN 1978K (180'' $\times$ 180'' section).
The source has an extremely red color
with a flux ratio of
$F_{\nu}$(S7)/$F_{\nu}$(S11) = 0.1,
which is out of a typical range in 
the spiral arm and the inter-arm of the galaxy
($\sim 0.8-1.4$, \citealt{sakon07}).
The origin of this emission is discussed in 
Section \ref{sec:1978K}.

For the other 5 objects, we do not detect 
a significant signal from SNe.
Upper limits of the flux are shown in Table \ref{tab:phot}.

\section{Dust Models}
\label{sec:model}

In the following sections,
we compare the observations with various model spectra.
We consider two types of carbon dust grains, \ie 
amorphous carbon and graphite,
and two kinds of silicate dust grains, \ie  
``astronomical silicate'' and forsterite (${\rm Mg_2SiO_4}$).
We assume that grains are spherical.

Thermal emission flux from the dust toward a SN can be expressed as follows;
\begin{equation}
F_{\lambda} = \Md \left( \frac{4\pi}{3}\rho a^3 \right)^{-1}
4\pi a^2 Q_{\lambda}^{\rm abs} \pi B_{\lambda}(\Td) 
\left( \frac{1}{4 \pi d^2} \right),
\label{eq:flux}
\end{equation}
where $\Md$ is the total dust mass, 
$\rho$ and $a$ are, respectively, 
the density and the radius of the dust grain, 
$Q_{\lambda}^{\rm abs}$ is the absorption efficiency factor,
$\Td$ is the temperature of the dust, and $d$ is the distance to the SN.
We assume a single temperature for simplicity.
In the equation above, 
all of the IR emission from dust grains is assumed 
to be detected without being absorbed.

We consider a single size of dust $a=0.1$ \mum.
In reality, there is a distribution in the grain size,
and if dust grains are swept up by the shock waves of the
SN, their size distribution can change with time \citep{nozawa07}.
However, $Q_{\lambda}^{\rm abs}/a$ is 
independent of the grain size if $\lambda \gg 2 \pi a$, 
and thus, the assumption of the single-size dust does not 
affect the estimate of the dust mass.

For the grain densities, we adopt $\rho =$ 2.26, 2.26, 3.8, and 3.2
${\rm g\ cm^{-3}}$
for amorphous carbon, graphite, astronomical silicate, and forsterite,
respectively.
The absorption efficiency factor is taken from 
\citet{edoh83} for amorphous carbon,
\citet{draine03Si} for graphite and astronomical silicate,
and \citet{semenov03} for forsterite.

Since the observed flux in Table \ref{tab:phot} is {\it not}
color-corrected due to the unknown intrinsic 
spectrum of the sources,
we apply the color corrections to the model.
Since the flux density of the AKARI/IRC data refers to the spectrum 
$F_{\lambda} \propto 1/\lambda$,
the following correction is applied to the model spectrum 
$F_{\lambda}^{\rm model} (\lambda)$;
\begin{eqnarray}
F_{\lambda}^{\rm IRC} (\lambda_i)
= \frac{\int F_{\lambda}^{\rm model}(\lambda) R (\lambda) d\lambda}
{\int (\lambda_i/\lambda) R(\lambda) d\lambda}.
\end{eqnarray}
Here, $\lambda_i$ is the reference wavelength of the band,
and $R (\lambda)$ is the response function 
(electron energy$^{-1}$).
We use the response function measured in the laboratory 
\footnote{http://www.ir.isas.jaxa.jp/AKARI/Observation/}
\citep[see][]{onaka07,lorente07}.
The calibration of the Spitzer/MIPS data uses
the blackbody spectrum with $T_0 = 10000$ K as the reference.
Thus, we apply the following correction;
\begin{eqnarray}
F_{\lambda}^{\rm MIPS} (\lambda_i)
= \frac{\int F_{\lambda}^{\rm model}(\lambda) R (\lambda) d\lambda}
{\int  \left( \displaystyle \frac{\lambda_i}{\lambda} \right)^5  
\displaystyle \frac{{\rm exp}({hc/\lambda_i k T_0}) - 1}{{\rm exp}({hc/\lambda k T_0}) - 1} 
R(\lambda) d\lambda}.
\end{eqnarray}
We use the response function provided by the Spitzer Science Center
\footnote{http://ssc.spitzer.caltech.edu/mips/calibrationfiles/}.

\begin{figure}
\begin{center}
\includegraphics[scale=1.3]{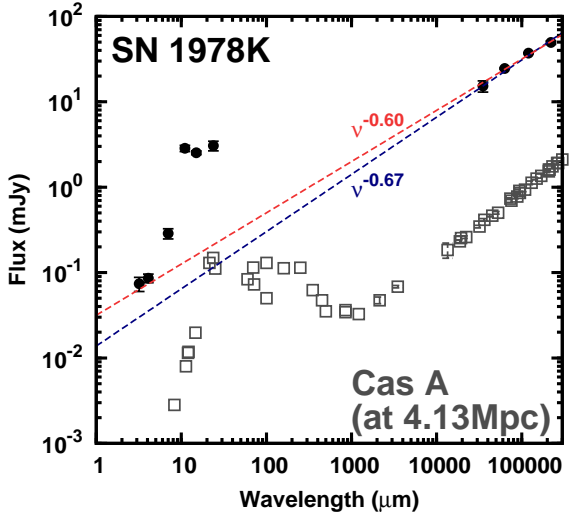}
\caption{
The IR-radio SED of SN 1978K (filled circles) compared with
that of Cas A (open squares) at the same distance (4.13 Mpc).
The red dashed line shows the power-law fit
to the IR and radio data of SN 1978K while the blue dashed line
shows the fit only to the radio data.
The radio data of SN 1978K are taken from \citet{smith07}.
The data of Cas A are taken from 
\citet{baars77}, \citet{mezger86}, \citet{liszt99},
\citet{hines04}, and \citet{barlow10}.
\label{fig:flux78K}}
\end{center}
\end{figure}

\begin{figure}
\begin{center}
\includegraphics[scale=1.3]{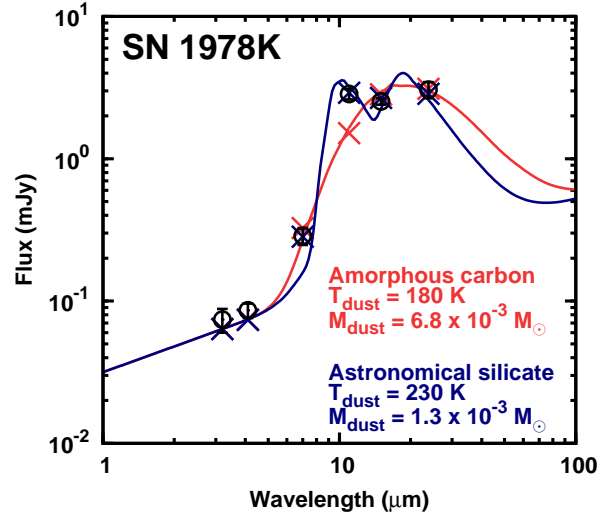}
\caption{
The IR SED of SN 1978K (open black circle) compared with
the model spectra.
The red line shows the spectrum of the amorphous carbon with 
$\Td = 180$ K and $\Md = 6.8 \times 10^{-3} \Msun$
while the blue line shows the spectrum of astronomical silicate with 
$\Td = 230$ K and $\Md = 1.3 \times 10^{-3} \Msun$.
The crosses show the color-corrected model flux for each band.
For both models, the synchrotron component 
($F_{\nu} \propto \nu^{-0.60}$) is added.
In the case of astronomical silicate, the color corrections are
$F_{\nu}^{\rm IRC, MIPS}(\lambda_i) / F_{\nu}^{\rm model}(\lambda_i)$ = 
1.7, 0.91, 1.1, and 1.1
for the S7, S11, L15, and MIPS 24 \mum\ band, respectively.
In the case of amorphous carbon, these are 
1.1, 0.93, 0.95, and 1.0, respectively.
\label{fig:model78K}}
\end{center}
\end{figure}

\section{IR Emission from SN 1978K}
\label{sec:1978K}

\subsection{Spectral Fit}
\label{sec:fit}

Among our 7 targets, we detect IR emission only from SN 1978K.
SN 1978K is an extraordinarily strong Type IIn SN, 
which is visible both in radio and X-ray even at
$\sim 30$ yr after the explosion 
\citep{ryder93,petre94,chugai95,schlegel96,montes97,schlegel99,chu99,
schlegel00,schlegel04,lenz07,smith07}.
Such long-lasting radio and X-ray emissions are 
clear evidence for the presence of the dense circumstellar medium, 
with which the SN ejecta has been interacting 
\citep{chevalier82,chevalier94,chugai95}.

Figure \ref{fig:flux78K} shows the spectral
energy distribution (SED) of SN 1978K from IR to 
radio wavelengths (filled circles).
The SED of Cassiopeia A (Cas A) at the distance 
of NGC 1313 is also shown for comparison (open squares).
These two objects show similar overall SEDs from
IR to radio wavelengths.

Since SN 1978K is powerful in radio wavelengths
\citep{smith07}, we must take into account the 
influence of the synchrotron emission on the IR bands
\citep[see \eg][for the case of Cas A]{mezger86, rho03, dwek04,
hines04, barlow10}.
The radio emission of SN 1978K is well fitted by a power-law spectrum 
$F_{\nu} \propto \nu^{\alpha}$ \citep{montes97,smith07}.
When we fit the flux of 4 radio bands of SN 1978K
\citep{smith07},
we derive $\alpha = -0.67 \pm 0.05$ 
(blue dashed line in Figure \ref{fig:flux78K}).
The flux densities at the N3 and N4 bands 
are almost in line with the power-law spectrum.
If we include the flux of these 2 bands for the fitting,
we obtain $\alpha = -0.60 \pm 0.02$ (red dashed line).
Since these two values are consistent within the error,
we assume that the emissions at the N3 and N4 bands are of
synchrotron origin
and use $\alpha = -0.60$ in the following discussions. 

The flux density of the S7, S11, L15, and L24 (MIPS 24 \mum) bands 
is clearly above the synchrotron component.
Therefore, we conclude that this excess is caused by 
the dust emission associated with SN 1978K.

\begin{figure*}
\begin{center}
\begin{tabular}{cc}
\includegraphics[scale=1.1]{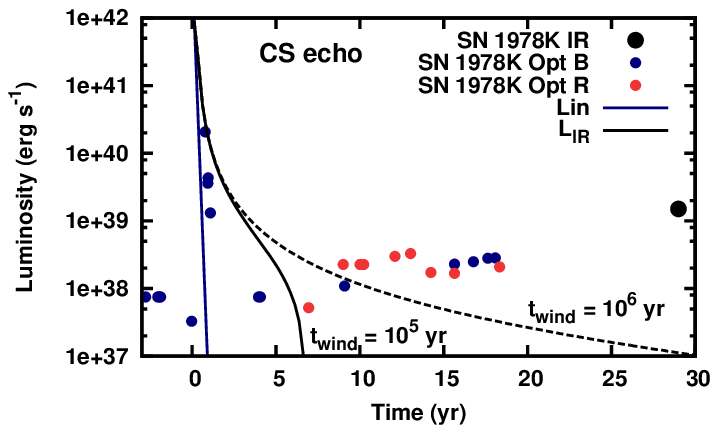} &
\includegraphics[scale=1.1]{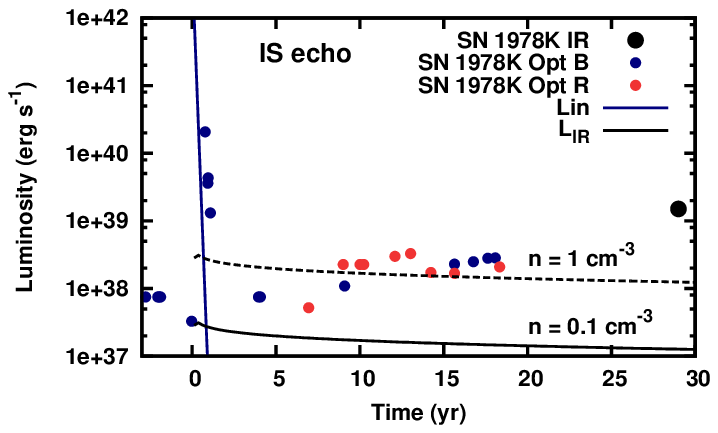} 
\end{tabular}
\caption{
(Left) Computed light curves of the IR echo by the CS dust (lines),
compared with observations of SN 1978K (points).
The purple line shows the input SN luminosity, while 
the black line shows the IR luminosity by the light echo.
The optical data points are shown by assuming zero 
bolometric correction using the data by \citet{ryder93,schlegel99}.
The IR echo of the SN outburst luminosity
$L = L_0 \exp(t/t_{\rm SN})$.
The solid line shows the model with 
$L_0 = 6 \times 10^{42}\ {\rm erg\ s^{-1}}$,
$t_{\rm SN} = 25$ days, $\Mdot = 10^{-4}\ \Msun\ {\rm yr^{-1}}$, 
and $t_{\rm wind} = 10^5$ yr.
Dashed lines show the extreme cases with $t_{\rm wind} = 10^6$ yr. 
Even with such an extreme assumption, the expected luminosity 
is lower than the observed IR luminosity at $t=30$ yr.
(Right) The same with the left panel but for the IS dust echo.
Black solid and dashed line shows the IR luminosity for 
$n_0 = 0.1$ and $1\ {\rm cm^{-3}}$, respectively. 
\label{fig:echo}}
\end{center}
\end{figure*}

We fit the observed flux with the model spectra
of the 4 dust species with 
the dust temperature and mass as parameters. 
The quality of the fit is judged by $\chi^2$ for
the 4 observational points, \ie the IRC S7, S11, L15, and MIPS 24 \mum\ bands.
For each dust species, we choose the best model 
that gives the smallest $\chi^2$.
Since the dependence on the temperature and mass is simple, 
the fit is unique, \ie there is no other solution 
with a different set of temperature and mass.

Among the 4 dust species, the smallest $\chi^2$ is 
obtained for astronomical silicate with $\Td = 230$ K and
$\Md =1.3 \times 10^{-3} \Msun$ 
(blue line in Figure \ref{fig:model78K}).
The total IR flux by the dust emission is 
$F_{\rm IR} = 7.0 \times 10^{-13}\ {\rm erg\ cm^{-2}\ s^{-1}}$
and the total IR luminosity is 
$L_{\rm IR} = 1.5 \times 10^{39} {\rm erg\ s^{-1}}$.
Silicate dust grains are favored over carbon grains because 
the observation shows excess at the S11 band,
which can be attributed to the Si-O stretching mode of 
silicate dust at 10 \mum\ \citep{draine84,draine03}.
As shown in the red line in Figure \ref{fig:model78K},
amorphous carbon dust grains are not able to explain the excess.
Forsterite fits the observations quite well, 
but gives a band feature at 10 \mum\ slightly narrower than
astronomical silicate.
Hereafter, we simply call the dust species found in 
SN 1978K ``silicate dust''.

\subsection{Origin of the IR emission}

SN 1978K is associated with 
$1.3 \times 10^{-3} \Msun$ of silicate dust.
The derived temperature is relatively high (230 K).
In this subsection, we discuss possible origins of 
the IR emission.
Since SN 1978K is a CSM-rich SN, we first consider CS dust.
CS dust can radiate IR emission by light echo (Section \ref{sec:echo}).
Then, the echoing CS dust is located at the distance larger than
$D = ct/2 =  1.4 \times 10^{19} (t/30\ {\rm yr})$ cm,
where the SN forward shock cannot reach.
On the other hand, the CS dust swept up by the SN forward shock 
can be heated via the collisions with energetic electrons 
and emits thermal radiation (Section \ref{sec:CSdust}).
Then, the typical location is at $\sim 5 \times 10^{17} 
(t/30\ {\rm yr})(\vsn/5000\ {\rm km\ s^{-1}})$ cm.
Another possibility is the SN dust (Section \ref{sec:SNdust}).
We show that, among these possibilities, 
the shocked CS dust is the most probable origin for 
the IR emission from SN 1978K.

\subsubsection{Light Echo}
\label{sec:echo}

The first possibility for the IR emission is 
echo emission of CS dust.
The dust can absorb the ultra-violet or optical 
photons of a SN, and radiate IR emission
\citep{bode80,wright80,dwek83,dwek85,graham86}.
In fact, the presence of CS dust echo 
has been suggested by early phase observations of SNe
\citep{meikle06,meikle07,andrews11}
and SN 1980K at the transitional phase \citep{sugerman12}.

We calculate the luminosity of the light echo 
by following \citet{dwek83}.
The gas mass-loss rate of the progenitor of SN 1978K is
estimated to be about $\Mdot = 10^{-4}\ M_\sun\ {\rm yr^{-1}}$
from optical, X-ray and radio observations 
\citep{ryder93,chugai95,smith07}. 
The duration of mass loss is assumed to be $t_{\rm wind} = 10^{5}$ yr.
This is a rather extreme assumption since the 
progenitor loses mass as large as 10 $\Msun$ prior to the explosion.
We assume that the wind velocity is $\vwind=10$ \kms,
which gives outer radius of CSM about $R_2= 3 \times 10^{18}$ cm.
The mass-loss rate is assumed to be constant, 
and thus, the CSM density has a power-law radial profile $\propto$ $r^{-2}$.

For the SN luminosity, we use an exponential form
$L = L_0 \exp(t/t_{\rm SN})$.
We assume a normal peak luminosity 
$L_0 = 6 \times 10^{42}\ {\rm erg\ s^{-1}}$ 
and $t_{\rm SN} = 25$ days
\citep{sahu06,kotak09}.
The other parameters are set as follows;
the gas-to-dust mass ratio $f=200$, 
the grain radius $a=0.1$ \mum,
the mean absorption efficiency $\bar{Q} = 1$,
and the evaporation temperature of dust $T=1500$ K.
The SN outburst luminosity will make a dust cavity with a
radius of $R_1 = 7 \times 10^{16}$ cm.

The left panel of Figure \ref{fig:echo} shows the computed IR luminosity 
of the CS dust echo (solid black line).
Optical data of SN 1978K are also plotted for comparison
under a simple assumption of zero bolometric correction (blue and red points).
The IR echo luminosity at $t=10-100$ yr
depends largely on the outer radius ($R_2$) of the CSM
since the luminosity from the CS echo declines
rapidly at $t = 2R_2/c \sim 7$ yr \citep{dwek83}.
To illustrate this dependence, we also show the model with 
a longer mass-loss duration (dashed line for $t_{\rm wind}=10^6$ yr).
With $t_{\rm wind} \gsim 5 \times 10^{5}$ yr, with which
the progenitor loses more than 50 $\Msun$, 
the IR echo does not drop dramatically at $t<$30 yr.
But, even with this extreme case,
the IR luminosity is only an order of $10^{37}\ {\rm erg\ s^{-1}}$
at $t=30$ yr, which is much less than the observed IR luminosity.
Thus, we conclude that the CS dust echo of SN outburst is unlikely
to be the origin of the IR emission.

We also calculate the IS dust echo by the same method
as that for the CS dust echo.
In fact, IR echo by the IS dust has been detected
for several extragalactic SNe \citep{meikle07,kotak09,meikle11}
and Cas A \citep{krause05,dwek08b}.
The difference from the CS dust is that the density is set 
to be uniform and that the much larger outer radius is allowed. 
Here,
the distribution of IS dust is approximated 
to be a sphere with the radius of $R_2 =$ 100 pc 
($t = 2R_2/c \sim 700$ yr).
The black solid and dashed lines in the right panel of 
Figure \ref{fig:echo} show the luminosity evolution of the IS dust echo
for the ISM gas density of $n = 0.1$ and $1\ {\rm cm^{-3}}$, respectively.

With the normal SN luminosity, the expected IR luminosity at $t=30$ yr
is not high enough to explain the observed luminosity of SN 1978K.
To explain the observation, the SN had to be very luminous, 
$L_0 \sim 6 \times 10^{43}\ {\rm erg\ s^{-1}}$, 
or the absolute optical magnitude of about $-20$ mag.
It would imply that the brightness of SN 1978K was of the 8th magnitude.
This is in contrast to the suggestion by \citet{ryder93} 
that SN 1978K was subluminous.
Although a very large luminosity of SN 1978K is not completely 
excluded given the sparse observations in 1978,
we conclude that IS dust echo is also unlikely.

\begin{figure}
\begin{center}
\includegraphics[scale=1.2]{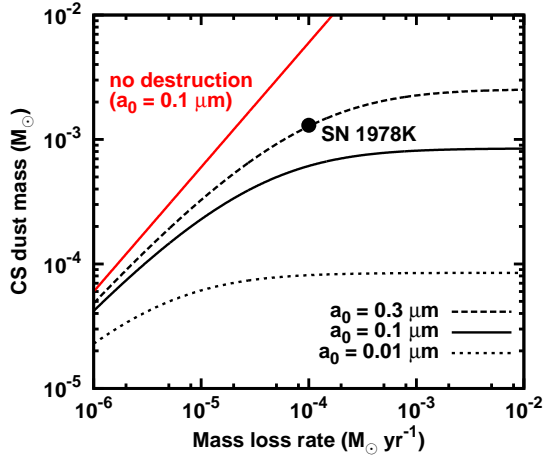} 
\caption{
The mass of shocked CS dust that survives 
at $t=30$ yr after the explosion as a function of mass-loss rate.
The dashed, solid and dotted lines shows the model 
with $a_0 = 0.3, 0.1$, and $0.01$ \mum, respectively.
The red dashed line shows the model without dust 
destruction with $a_0 = 0.1$ \mum.
The other parameters are set as follows;
$\vsn = 4000$ \kms, $\vwind = 10$ \kms, and $f=200$.
\label{fig:CSshock}}
\end{center}
\end{figure}

\subsubsection{Shocked CS dust}
\label{sec:CSdust}

The second source for the IR emission 
is the CS dust heated by the forward shock of the SN.
The mid-IR emission from the ring in SN 1987A 
is thought to arise from this mechanism \citep[\eg][]{dwek08}.
It is interesting to note that 
the emission properties of SN 1987A ring and SN 1978K look similar.
The dust temperature in the SN 1987A ring is 
$\sim 165-185$ K at $t=$ 17.9 -- 19.7 yr \citep{bouchet06,dwek08,seok08},
while that in SN 1978K is 230 K at $t=$ 29 yr.
In addition, the IR to X-ray flux ratio is of the order 
of unity in both objects 
(For SN 1978K, $F_{\rm X} \sim 1 \times 10^{-12}\ {\rm erg\ cm^{-2}\ s^{-1}}$, 
\citealt{smith07}).
The similarity seems to support the shock-heated CS dust
as an origin of IR emission from SN 1978K.

We estimate the mass of the emitting CS dust 
swept by the forward shock.
CS dust is subject to the destruction by 
the SN forward shock.
The timescale of sputtering is given by 
\begin{equation}
\tau_{\rm sput} \equiv
\left( \frac{1}{a}\frac{da}{dt}  \right)^{-1}
\simeq 10^6 \ {\rm yr} \
\left( \frac{a}{1 \mu {\rm m}} \right)
\left( \frac{n_{\rm g}}{1 \ {\rm cm^{-3}}} \right)^{-1}
\end{equation}
\citep{dwek92araa}, 
where $n_{\rm g}$ is the number density of shocked gas.
Then, the evolution of the dust radius is estimated as 
$a(t) = a_0 (1 - t/\tau_{\rm sput}^{\prime})$,
where $\tau_{\rm sput}^{\prime} = 10^6\ {\rm yr} (a_0/1 \mu {\rm m}) 
(n_{\rm g}/ 1 \ {\rm cm^{-3}})^{-1}$.
Hereafter, the subscript 0 denotes the value before the dust destruction.
The gas mass-loss rate and the CS dust density is related as follows;
$\dot{M} = 4 \pi R^2 \rho_{\rm wind} \vwind 
= 4 \pi R^2 f \rho_{\rm dust, 0} \vwind$,
where $\rho_{\rm wind}$ is the mass density of the wind gas.
We also define the mass density of dust before the destruction;
$\rho_{\rm dust, 0} \equiv (4/3) \pi \rho a_0^3 n_{\rm dust, 0}$,
where $n_{\rm dust, 0}$ is the number density of dust before the 
destruction.
The CS dust density after the destruction is given by
$\rho_{\rm dust}(t) = (4/3) \pi n_{\rm dust, 0} \rho a^3 
= (4/3) \pi n_{\rm dust, 0} \rho a_0^3 
(1 - t/\tau_{\rm sput}^{\prime})^3 
= \rho_{\rm dust, 0} (1 - t/\tau_{\rm sput}^{\prime})^3$.

Using these relations, 
the mass of the emitting, shocked CS dust is described
as follows;
\begin{eqnarray}
&& M_{\rm CS dust} (t) =
\int^{t}_{0} 4 \pi R^2 \rho_{\rm dust}(t^{\prime}) \vsn dt^{\prime} \nonumber \\
=&& \int^{t}_{0} 4 \pi R^2 \rho_{\rm dust, 0} \vsn 
(1 - t^{\prime} /\tau_{\rm sput})^3 dt^{\prime} \nonumber \\
=&& 8 \times 10^{-4} \Msun
\left( \frac{\dot{M}_{\rm wind}}{10^{-5}\ \Msun\ {\rm yr^{-1}}} \right) 
\left( \frac{\vsn}{5000\ {\rm km\ s^{-1}}} \right) \nonumber \\ 
\times && 
\left( \frac{\vwind}{10\ {\rm km\ s^{-1}}} \right)^{-1}
\left( \frac{f}{200} \right)^{-1} 
\left( \frac{\int^t_0 (1 - t^{\prime} /\tau_{\rm sput}^{\prime})^3 dt^{\prime}}{30\ {\rm yr}}
\right).
\end{eqnarray}
At the limit of no dust destruction, 
the integral in the last parentheses becomes $t$,
and the shock-heated CS dust is proportional to the mass-loss rate.
But $\tau_{\rm sput}^{\prime}$ is a function of the shocked gas density,
and thus, the mass-loss rate.
With a larger mass-loss rate, the sputtering timescale becomes shorter,
and the integral in the last parentheses becomes smaller than $t$.

Figure \ref{fig:CSshock} shows the shocked (but surviving) mass of 
the CS dust at $t=30$ yr as a function of mass-loss rate.
The dashed, solid and dotted lines in black show the models with
$a_0 = 0.3, 0.1$, and $0.01$ \mum, respectively.
The red solid line is the model without dust destruction 
with $a_0 = 0.1$ \mum.
Other parameters are assumed as follows;
$\vsn = 4000$ \kms, $\vwind = 10$ \kms, and $f=200$.
The SN shock velocity of SN 1978K is estimated to be $\vsn < 4000$ \kms
by the VLBI image, which marginally resolve it \citep{smith07}.
Thus we assume $\vsn = 4000$ \kms.

For small mass-loss rates, the destruction is not so effective
that the shocked CS dust mass is roughly proportional to the mass-loss rate.
For large mass-loss rates, the surviving dust mass is not 
sensitive to the mass-loss rate.
Because of the high gas density, 
the destruction is effective enough 
that the mass of dust swept up by the shock compensates for
the mass destroyed by the shock
\footnote{\citet{fox10,fox11} also derived the expression for the
shock-heated dust mass, which is independent on the mass-loss rate.
Since they assume a destruction-free shell,
the mass of CS dust is larger than our calculation by 
a factor of 4.}. 
In the calculation above, we assume there is no dust-free cavity 
created by the SN outburst.
Since most of surviving dust is located near the 
SN shock (at $R \sim 5 \times 10^{17} 
(t/30\ {\rm yr})(\vsn/5000\ {\rm km\ s^{-1}})$ cm), 
which is larger than the cavity size
($\sim 7 \times 10^{16}$ cm).
Thus, the results presented here are not largely affected by
the presence of the dust-free cavity.

From the IR SED of SN 1978K, 
we estimate the mass of emitting dust to be $1.3 \times 10^{-3} \Msun$.
The model of the shocked CS dust 
with $\dot{M} = 10^{-4}\ \Msun\ {\rm yr^{-1}}$ and $\vsn = 4000$ \kms\
is consistent with the observed dust mass 
if the initial dust radius is $a_0 =0.3$ \mum.
A similar conclusion has been reached by \citet{dwek08} for SN 1987A.

We also check the expected temperature for the shocked CS dust.
With $\Mdot = 10^{-4}\ \Msun\ {\rm yr^{-1}}$, 
the shocked gas density is about $6 \times 10^{3}\ {\rm cm^{-3}}$.
For the electron temperature about $10^7$ K, 
the expected dust temperature is $\Td \sim 270$ and 220 K for 
the dust radius $a = 0.1$ and 0.3 \mum, respectively \citep{dwek08}.
Thus, the expected temperature of the shocked CS dust is 
also in good agreement with the observations.
The surviving, shocked CS dust is distributed in a thin shell,
whose inner radius is about 95 \% of the radius at the forward shock. 
Using the mass absorption coefficient of dust $\kappa_{\nu}$
for astronomical silicate at 24 \mum\ \citep{draine03Si},
the optical depth of dust is about 
$4 \times 10^{-4} (\kappa_{\nu}/562 \ {\rm cm^2\ g^{-1}})$,
which is consistent with our assumption of optically thin dust
(Section \ref{sec:model}).
In summary, the IR emission from the shocked CS dust 
can naturally explain the observed IR emission from SN 1978K.

\subsubsection{SN dust}
\label{sec:SNdust}

The last possibility for the origin of the IR emission is 
dust formed in SN ejecta.
Theoretically, SNe are expected to form dust masses as large as 
$\Md = $ 0.1$-$1.0 $\Msun$ at the inner layer of the SN ejecta
\citep{kozasa89,kozasa91,todini01,nozawa03,nozawa08,nozawa10,
bianchi07,chercheneff10}.
In fact, such a large amount of dust is required 
to explain the observed dust mass in high-redshift quasars.
if the dust is formed only by SNe
\citep{dwek07,meikle07,gall11}.

Observationally the mass
of dust formed by SNe is quite controversial.
IR observations of young SNe ($t \lsim 1-2$ yr) 
in external galaxies
usually find only $10^{-5} - 10^{-3} \Msun$ of dust
\citep[\eg][]{wooden93,ercolano07,elmhamdi03,meikle07,kotak09,szalai10,
meikle11}.
On the other hand, observations of supernova remnants (SNRs)
in our Galaxy and Large/Small Magellanic Clouds
detect a larger amount of dust.
For example, by intensive observations of Cas A
at IR to sub-mm wavelengths, it is estimated 
that $\sim 0.1 \Msun$ of dust is formed
in total \citep{rho08,sibthorpe10,barlow10}.
Recently, \citet{matsuura11} detected a large amount of 
cold dust in SN 1987A at $t=23.3$ yr 
in far-IR and sub-mm wavelengths.
The estimated dust mass is as large as $\sim 0.4-0.7 \Msun$,
and the dust temperature is $\sim 20$ K.

To estimate the IR emission from SN dust, 
we first consider the total luminosity available, which
gives an upper limit of the IR luminosity.
At the transitional phase, one of the major heating sources
of SN dust formed at the inner layers of the ejecta
is the radioactive decay of $^{44}$Ti, which gives 
$L \sim 10^{36} - 10^{37} {\rm erg\ s^{-1}}$
\citep{fransson02}.
This is much smaller than the IR luminosity 
($L_{\rm IR} \sim 1.5 \times 10^{39} {\rm erg\ s^{-1}}$) of SN 1978K.

A possible additional heating source is UV or X-ray emission 
from the ejecta-CSM interacting region.
X-ray luminosity of SN 1978K is very high
($3 \times 10^{39} {\rm erg\ s^{-1}}$ \citep{smith07}, 
being enough to account for the observed IR luminosity.
However, the X-ray emission from the shocked region 
is absorbed by the optically-thick ejecta 
via photo-absorption even at the transitional phase \citep{fransson87},
and it cannot reach the inner layer 
and cannot heat the SN dust there effectively.
Therefore, SN dust at the inner layers of the ejecta 
is quite unlikely to be the origin of the IR emission from SN 1978K.

Note that dust formation in a cool dense shell has been proposed 
for some interacting SNe
\citep[\eg][]{pozzo04,smith08,mattila08,fox09,smith09,chugai09}.
The cool dense shell may be formed at the reverse-shocked region,
and thus, the dust can be heated by the UV/X-ray radiation.
Although it is not clear 
whether $1.3 \times 10^{-3} \Msun$ of the dust can be formed 
in the cool dense shell, 
this scenario cannot be completely excluded as a possible origin 
for the IR emission from SN 1978K.

\begin{figure}
\begin{center}
\includegraphics[scale=1.2]{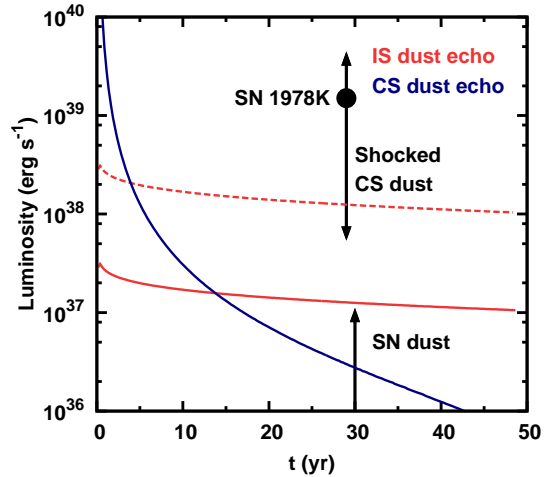}
\caption{
Expected IR luminosity for normal SNe.
The red and blue lines show the echo of the IS dust 
and the CS dust, respectively.
For the IS dust echo, the red solid and dashed 
lines show the case for $n=0.1$ and $1\ {\rm cm^{-3}}$, 
respectively.
For the CS dust echo, the solid line 
shows the case for
$\Mdot = 10^{-5} \ \Msun \ {\rm yr^{-1}}$ and
$t_{\rm wind} =$ $10^{6}$ yr 
(outer radius $R_2 = 3 \times 10^{19}$ cm).
The expected luminosities of the SN dust and 
the shocked CS dust are indicated with the arrows.
\label{fig:IRlum}}
\end{center}
\end{figure}

\section{IR Emission from Normal SNe}
\label{sec:normalSN}

From the upper limits of the flux for 5 non-detected SNe, 
upper limits of the emitting dust mass can be obtained.
In this section, we first summarize 
expected IR emission from normal SNe at the transitional phase.
Then, we discuss the implication of the non-detection
by current observations.

\begin{figure}
\begin{center}
\begin{tabular}{c}
\includegraphics[scale=1.2]{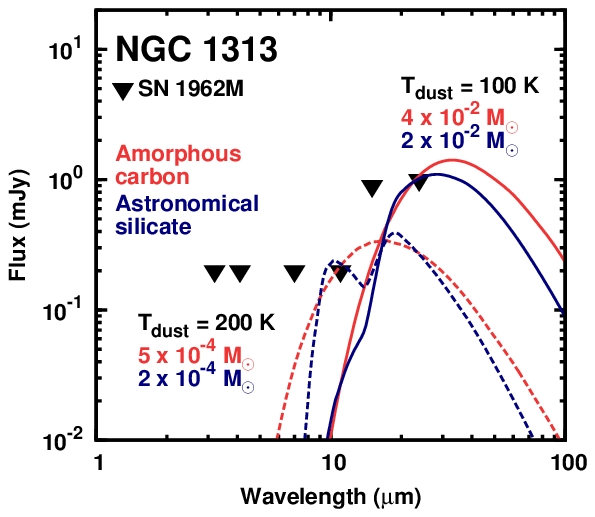} \\
\includegraphics[scale=1.2]{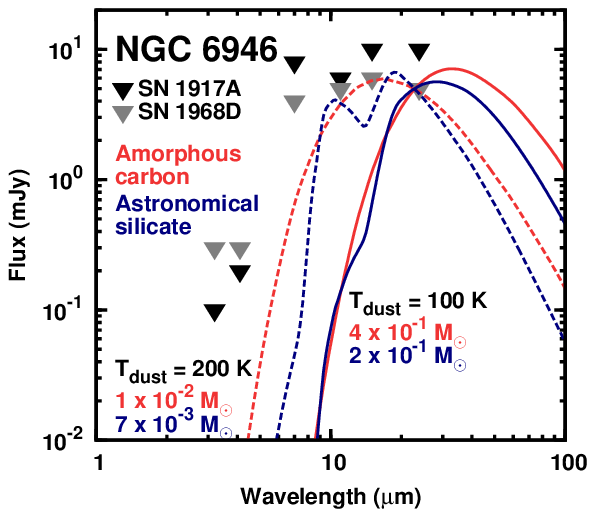} \\
\includegraphics[scale=1.2]{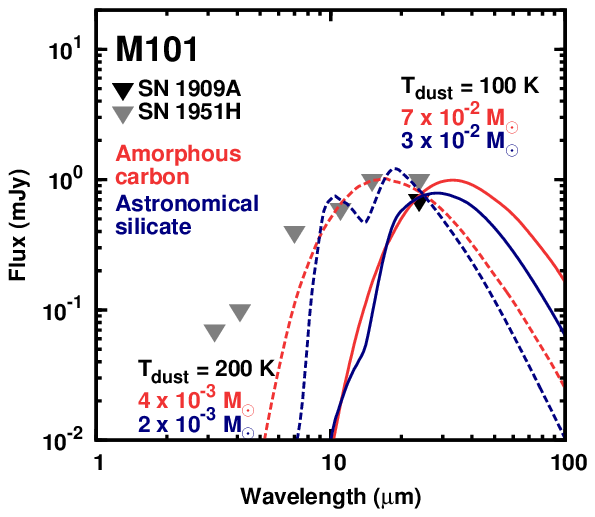}
\end{tabular}
\caption{
The upper limits of the flux for 6 non-detected SNe 
in NGC 1313, NGC 6946, and M101 (from top to bottom).
The lines show the model spectra
of amorphous carbon (red) and astronomical silicate (blue)
with $\Td$ = 200 K (dashed) and 100 K (solid).
The dust mass assumed in the model is shown in the panels.
\label{fig:sed}}
\end{center}
\end{figure}

\subsection{Summary of IR Emission at the Transitional Phase}
\label{sec:expectation}

For CSM-rich SN 1978K, we have discussed IR emission 
expected by (1) light echo from CS and IS dust, 
(2) shocked CS dust, and (3) SN dust.
In this section, we summarize the  
IR emissions from normal SNe by these processes, 
presenting the expected IR luminosity and typical dust temperature.

We have shown the light curve of the light echo 
by CS dust for $\Mdot = 10^{-4}\ \Msun\ {\rm yr^{-1}}$
in Figure \ref{fig:echo}.
Using the same method, we calculate the 
echo luminosity with a normal mass-loss rate
$\Mdot = 10^{-5}\ \Msun\ {\rm yr^{-1}}$, 
and a normal SN luminosity
$L_0 \sim 6 \times 10^{42}\ {\rm erg\ s^{-1}}$.
The solid blue line in Figure \ref{fig:IRlum} shows
the IR luminosity with the wind duration $t_{\rm wind} = 10^6$ yr.
The echo luminosity is about $L = 3 \times 10^{36}\ {\rm erg\ s^{-1}}$ 
at $t=30$ yr.

Figure \ref{fig:IRlum} also shows the 
luminosity by IS dust echo as calculated in Section \ref{sec:echo}.
The light curve of the IS dust echo has a long timescale, 
and will exceed that of the CS dust echo after $t=5-10$ yr
for the ISM density of $n = 0.1$--1.0 cm$^{-3}$.
Note that a typical interstellar density is not well constrained. 
\citet{gaustad93} derived a filling factor of 
interstellar gas with density $> 0.5\ {\rm cm^{-3}}$
to be only about 15 \%.
\citet{dring96} derived an even smaller filling factor:
8.5 \% for the gas density $> 0.01 \ {\rm cm^{-3}}$.
When the density is lower than 0.1 ${\rm cm^{-3}}$,
the IR luminosity by the echo is roughly proportional 
to the gas density since the IS dust is optically thin.

For both CS and IS dust cases, the typical 
temperature of IR emission can be roughly estimated 
at the closest point from the SN,
\ie at $D = ct/2 = 1.4 \times 10^{19} (t/30\ {\rm yr})$ cm,
which has the largest contribution to the total luminosity.
Suppose that a dust grain is irradiated by a SN 
with the luminosity $L_{\rm SN}$ at a distance $D$.
For simplicity, the spectrum of the luminous source 
is assumed to be blackbody with a temperature $T_s$.
Then, the absorbed luminosity is given by
$L_{\rm abs} = L_{\rm SN}/(4\pi D^2) \pi a^2 <Q^{\rm abs}(T_s)>$,
where $<Q^{\rm abs}(T)>$ is the Planck mean of the 
absorption efficiency factor.
When the dust particle emits radiation with a temperature $\Td$, 
the emitted luminosity is
$L_{\rm rad} = 4 \pi a^2 \sigma \Td^4  <Q^{\rm abs}(\Td)>.$
Under the equilibrium, \ie $L_{\rm abs} = L_{\rm rad}$, 
we can derive the temperature as follows;
\begin{eqnarray}
\Td &\simeq& 140\ {\rm K} 
\left( \frac{D}{1.4 \times 10^{19}\ {\rm cm}} \right)^{-1/3}
\left( \frac{L_{\rm SN}}{6 \times 10^{42}\ {\rm erg\ s^{-1}}} \right)^{1/6}  \nonumber \\
&& \times
\left( \frac{<Q^{\rm abs}(T_s)>}{1} \right)^{1/6}
\left( \frac{a}{0.1\ {\rm \mu m}} \right)^{-1/6}.
\label{eq:Tdust}
\end{eqnarray}
Here we used $<Q^{\rm abs}(\Td)>/a = 0.127 \Td^2$,
which is derived for $T \lsim 100$ K by \citet{dwek87}.
We find that this formula is applicable up to $T \sim 250$ K
for our purpose.
The expected temperature for the light echo is $\sim 140$ K
at $t=$30 yr. 

Next we consider the shocked CS dust.
Figure \ref{fig:CSshock} shows that the typical 
mass of the surviving, shock-heated dust 
is about $3 \times 10^{-4} \Msun$ at $t=30$ yr
for $\Mdot = 10^{-5}\ \Msun\ {\rm yr^{-1}}$
and $\vsn = 5000$ \kms.
With this mass-loss rate, the shocked gas density is 
about $6 \times 10^{2}\ {\rm cm^{-3}}$.
For the electron temperature about $10^7$ K, 
the expected dust temperature is $\Td \sim 160$ K 
for the grain radius of $a=0.1$ $\mu$m \citep{dwek08}.
The total IR luminosity can be derived 
by integrating Equation (\ref{eq:flux}) over wavelength;
\begin{eqnarray}
L_{\rm IR} &=&  \Md \left( \frac{4\pi}{3}\rho a^3 \right)^{-1}
4\pi a^2 \sigma \Td^4 <Q^{\rm abs}(\Td)>  \nonumber \\ 
&\simeq& 1.8 \times 10^{37}
\left( \frac{\Md}{0.1\ \Msun} \right)  \nonumber \\
&& \times
\left( \frac{\rho}{3.8\ {\rm g\ cm^{-3}}} \right)^{-1}
\left( \frac{\Td}{50\ {\rm K}} \right)^6\ \rm erg\ s^{-1},
\label{eq:LIR}
\end{eqnarray}
where, as in Equation (\ref{eq:Tdust}),
we adopt $<Q^{\rm abs}(\Td)>/a = 0.127 \Td^2$.
Using this equation, the total luminosity of the shocked CS dust is 
estimated as
$\sim 6 \times 10^{37}\ {\rm erg\ s^{-1}}$.
For a higher mass-loss rate $\Mdot = 10^{-4}\ \Msun\ {\rm yr^{-1}}$,
the shocked CS dust mass becomes $\Md \sim 1 \times 10^{-3} \Msun$, 
and the temperature is $\Td \sim 270$ K.
The total IR luminosity becomes $4 \times 10^{39}\ {\rm erg\ s^{-1}}$.

Finally, let us consider the SN dust.
For the SN dust, we assume 
$L_{\rm IR} < 10^{37}\ {\rm erg\ s^{-1}}$, 
which is expected by the radioactive decay of $^{44}$Ti.
If the mass of SN dust is $\Md =0.1\Msun$,
the temperature can be  $\Td <$ 50 K.
Even if only $\Md = 10^{-3} \Msun$ of dust is formed,
the temperature is about 100 K, which is lower than 
the expected temperatures for the light echo and the shocked CS dust.

In summary, IR emission of normal SNe at the transitional phase
can result from three types of mechanisms, \ie
IS dust echo, shocked CS dust, and SN dust.
IS dust echo and shocked CS dust have a relatively
high temperature ($\Td=150-270$ K)
while the SN dust has a lower temperature ($\Td < 50-100$ K).
The total luminosity of the IS echo can be 
$L_{\rm IR} > 10^{37}\ {\rm erg\ s^{-1}}$ 
for the IS density of $n > 0.1$ cm$^{-3}$
(Figure \ref{fig:IRlum}).
The shocked CS dust could also have 
$L_{\rm IR} \sim 6 \times 10^{37}- 4 \times 10^{39}\ {\rm erg\ s^{-1}}$
for $\Mdot =$ $10^{-5}$--$10^{-4}$ $\Msun\ {\rm yr^{-1}}$
(see the arrow in Figure \ref{fig:IRlum}).
The luminosity of the SN dust must be 
$L_{\rm IR} < 10^{37}\ {\rm erg\ s^{-1}}$.

\begin{deluxetable}{cccc} 
\tablewidth{0pt}
\tablecaption{Upper limit of Dust Mass}
\tablehead{
SN  & 
Age &
$\Md$ $(\Msun)$ &
$\Md$ $(\Msun)$ 
   \\
     & 
(yr) &
$\Td$=100 K &
$\Td$=200 K
}
\startdata
 &   & \multicolumn{2}{c}{Amorphous carbon} \\ \hline
SN 1909A    &  99   &  $<7 \times 10^{-2}$   &  $<3 \times 10^{-3}$  \\ 
SN 1917A    &  90   &  $<7 \times 10^{-1}$   &  $<3 \times 10^{-2}$  \\ 
SN 1951H    &  56   &  $<1 \times 10^{-1}$   &  $<4 \times 10^{-3}$  \\ 
SN 1962M    &  44   &  $<4 \times 10^{-2}$   &  $<5 \times 10^{-4}$   \\ 
SN 1968D    &  39   &  $<4 \times 10^{-1}$   &  $<1 \times 10^{-2}$  \\ \hline
 &   & \multicolumn{2}{c}{Astronomical silicate} \\ \hline
SN 1909A    &  99   &  $<3 \times 10^{-2}$   &  $<1 \times 10^{-3}$   \\ 
SN 1917A    &  90   &  $<3 \times 10^{-1}$   &  $<1 \times 10^{-2}$   \\ 
SN 1951H    &  56   &  $<5 \times 10^{-2}$   &  $<2 \times 10^{-3}$   \\ 
SN 1962M    &  44   &  $<2 \times 10^{-2}$   &  $<2 \times 10^{-4}$    \\ 
SN 1968D    &  39   &  $<2 \times 10^{-1}$   &  $<7 \times 10^{-3}$     \\ 
\enddata
\label{tab:upmass}
\end{deluxetable}

\begin{figure}
\begin{center}
\includegraphics[scale=1.3]{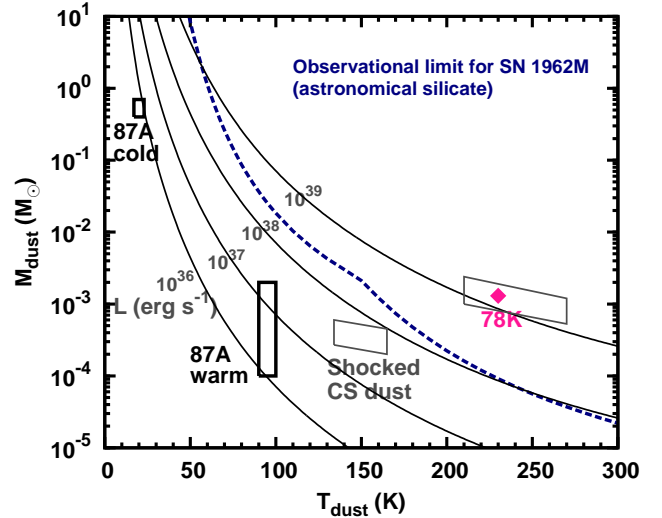}
\caption{
Upper limit of the dust mass as a function of the dust temperature
for the case of astronomical silicate.
The dashed blue line shows the upper limits for SN 1962M, 
which are derived from the non-detection in our current observation.
Solid curves show the upper limit for a given total luminosity 
$L = 10^{36} - 10^{39}\ {\rm erg\ s^{-1}}$.
Our current observations are sensitive to the luminosity
$L  \gsim 10^{38} {\rm erg\ s^{-1}}$.
The observations of SNe 1978K (pink diamond) and 1987A 
(black box, \citealt{bouchet04,matsuura11}) are also shown for comparison.
The expected ranges for the shocked CS dust with
$\Mdot =  10^{-5}\ \Msun\ {\rm yr^{-1}}$ (lower left) and 
$\Mdot =  10^{-4}\ \Msun\ {\rm yr^{-1}}$ (upper right) 
are shown (for dust size $a=0.1-0.3$ \mum).
\label{fig:mdust_sil}}
\end{center}
\end{figure}

\subsection{Constraints from Non Detections}

Figure \ref{fig:sed} shows SEDs of 5 SNe 
in the 3 different host galaxies.
These upper limits are compared with the model spectra
of amorphous carbon (red) and astronomical silicate (blue)
with $\Td$ = 100 and 200 K 
(solid and dashed lines, respectively).
The assumed dust mass is shown in the panels.
In Table \ref{tab:upmass}, we show the upper limit 
of the dust mass for the case of amorphous carbon 
and astronomical silicate for the temperature $\Td=$100 and 200 K.

The upper limit of the dust mass strongly depends 
on the dust temperature.
The dashed lines in Figure \ref{fig:mdust_sil} show
the upper limits of the dust mass for SN 1962M
as a function of the dust temperature
for the case of astronomical silicate
(blue and green dashed lines).
Constraints on the other 4 SNe are weaker 
than those for SN 1962M by a factor of 2--20.
For comparison, Figure \ref{fig:mdust_sil} also shows 
the estimated dust mass and temperature of SN 1978K (this paper) 
and both warm and cold component of SN 1987A \citep{bouchet04,matsuura11}.

The solid lines in Figure \ref{fig:mdust_sil} show
the dust mass as a function of dust temperature for 
$L_{\rm IR} = 10^{36}$, $10^{37}$, $10^{38}$, and 
$10^{39}\ {\rm erg\ s^{-1}}$ (from bottom to top).
This can be analytically derived from Equation (\ref{eq:LIR});
\begin{eqnarray}
\Md &=& 5.6 \times 10^{-2} 
\left( \frac{L_{\rm IR}}{10^{37}\ {\rm erg\ s^{-1}}} \right) \nonumber \\
&& \times \left( \frac{\Td}{50\ {\rm K}} \right)^{-6}
\left( \frac{\rho}{3.8\ {\rm g\ cm^{-3}}} \right)\ \Msun.
\label{eq:Mdust}
\end{eqnarray}
For these lines in Figure \ref{fig:mdust_sil},
we do not use the analytic formulae, but 
numerically integrate Equation (\ref{eq:flux}) over wavelength.

As shown in Figure \ref{fig:mdust_sil}, 
our current observations are sensitive only to the IR luminosity 
as high as $> 10^{38} {\rm erg\ s^{-1}}$.
This limiting luminosity is much higher than 
the predictions for the SN dust.
The luminosity of the IS dust echo 
can be $1 \times 10^{38}\ {\rm erg\ s^{-1}}$ with $1 \ {\rm cm^{-3}}$.
Our current observations can only exclude a homogeneous IS density 
of $n \sim 10 \ {\rm cm^{-3}}$.

For the shocked CS dust, 
the expected ranges of the dust mass and temperature
are shown by two gray boxes in Figure \ref{fig:mdust_sil}.
The left lower box shows the range for 
$\Mdot = 10^{-5}\ \Msun\ {\rm yr^{-1}}$,
while the right top for $\Mdot = 10^{-4}\ \Msun\ {\rm yr^{-1}}$,
Our current best observational limit excludes the 
existence of the shocked CS dust with
$\Mdot = 10^{-4}\ {\rm \Msun\ yr^{-1}}$.

\begin{figure*}
\begin{center}
\includegraphics[scale=1.4]{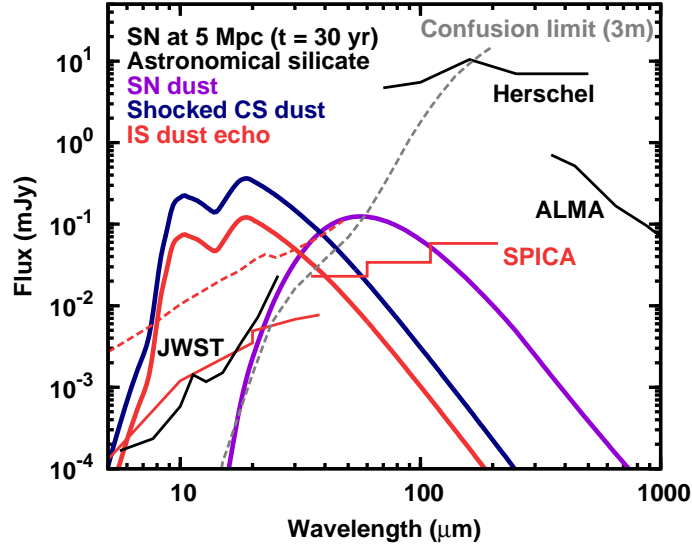}
\caption{
Expected IR SEDs of SN dust (purple), 
shocked CS dust (blue), and IS dust (red) compared with
$5\sigma$ detection limit of 1 hr imaging observation
with various telescopes, 
\ie JWST (black), SPICA (red), Herschel (black) and ALMA (black).
Gray dashed line shows the expected confusion limit of 
background galaxies with a 3 m telescope 
(T. T. Takeuchi et al. 2011, in preparation).
The distance to the SN is set to be 5 Mpc.
For the SN dust, we assume $\Md =0.1\Msun$
with $\Td$= 50 K.
This roughly corresponds to the total IR luminosity
of $L_{\rm IR} \sim 10^{37}\ {\rm erg\ s^{-1}}$.
For the shocked CS dust, we assume $\Md = 3 \times 10^{-4} \Msun$ 
with $\Td = 150$ K (the total luminosity is 
$L_{\rm IR} = 3.9 \times 10^{37}\ {\rm erg\ s^{-1}}$).
For the IS dust echo, we assume 
$L_{\rm IR} \sim 10^{37}\ {\rm erg\ s^{-1}}$
with $\Td = 150$ K. The effective emitting mass of 
dust is $\Md \sim 10^{-4} \Msun$.
\label{fig:SPICA}}
\end{center}
\end{figure*}

\section{Future Prospects}
\label{sec:future}

We show that our current observations are sensitive to 
the luminosity $L > 10^{38}\ {\rm erg\ s^{-1}}$.
In this section, we discuss prospects for future observations.
Figure \ref{fig:SPICA} shows 
the expected flux for the SN dust (purple,
$\Md = 0.1 \Msun$ and $\Td = 50$ K), 
the shock-heated CS dust (blue, 
$\Md = 3 \times 10^{-4}$ $\Msun$,
and $\Td = 150$ K, expected for $\Mdot = 10^{-5}\ \Msun\ {\rm yr^{-1}}$), 
and the IS dust echo (red, $L = 10^{37}\ {\rm erg\ s^{-1}}$
and $\Td = 150$K, the effective emitting dust mass is 
$\Md \sim 10^{-4} \Msun$).
The SN is assumed to be located at 5 Mpc.

The model spectra are compared with 
5$\sigma$ detection limit of 1 hr imaging observations
with various telescopes, \ie
with The James Webb Space Telescope 
(JWST, black, \citealt{wright04,swinyard04}), 
SPICA (red, \citealt{nakagawa10,wada10,swinyard08,kataza10}), 
Herschel (black, \citealt{pilbratt10,poglitsch10,griffin10}),
and ALMA (black, \citealt{wootten09}).
For SPICA, we also show the limit
for spectroscopic observations ($R=50$) 
in the red dashed line \citep{kataza10}.
In Appendix \ref{app:Mdust_all}
we show similar plots for other dust species.

The figure shows that 
future observations at $<30$\mum\ are sensitive to 
the shocked CS dust or IS dust echo.
These two cases may show similar overall spectra,
but ionic emission lines are also expected 
for the shocked CS dust.
Low resolution spectroscopy will be possible with SPICA, 
which will be useful to distinguish the shocked CS dust emission
from IS dust echo.

IR emission by SN dust has a smaller contribution 
to the IR luminosity.
However, thanks to the lower temperature than the 
shocked CS dust and IS dust echo, 
the expected spectrum has a peak at longer wavelengths.
The contamination by the shocked CS dust may be small
if $\Mdot \lsim 10^{-5}\ {\rm \Msun\ yr^{-1}}$.
In addition, the emission from the IS dust echo can 
also be small if a typical interstellar density is 
less than $0.1\ {\rm cm^{-3}}$ \citep{gaustad93,dring96}.
If the contributions from these two components are small, 
the detection of SN dust is limited only by the 
confusion limit of background galaxies, which is 
an order of 0.1 mJy at 50 \mum\ with a 3m telescope.

The great advantage to detect the SN dust at the transitional 
phase is the low optical depth.
If a SN forms 0.1 $\Msun$ of dust, the optical depth becomes
\begin{eqnarray}
\tau_{\nu} &=& \frac{3}{4\pi}\kappa_{\nu} \Md R^{-2}  \nonumber \\
&\simeq& 0.1 
\left( \frac{\kappa_{\nu}}{562\ {\rm cm^2\ g^{-1}}} \right)
\left( \frac{\Md}{0.1 \Msun} \right) \nonumber \\
&& \times
\left( \frac{\vsn}{5000\ {\rm km\ s^{-1}}} \right)^{-2}
\left( \frac{t}{30\ {\rm yr}} \right)^{-2}.
\label{eq:tau}
\end{eqnarray}
For the mass absorption coefficient of dust $\kappa_{\nu}$,
we take the value for astronomical silicate at 24 \mum\ \citep{draine03Si}.
At longer wavelengths, the coefficient is lower 
($\kappa_{\nu} \propto \lambda^{-2}$).
At early epochs, 
such a large mass of dust is optically thick even at IR wavelengths.
On the contrary, at the transitional phase,
the optical depth can be smaller than unity,
which enables a reliable estimate of the total dust mass.

Another advantage is the small confusion with 
the IS dust swept up by the SN ejecta.
The mass of the IS dust ($M_{\rm ISdust}$) 
can be approximately written as follows by neglecting the destruction;
\begin{eqnarray}
M_{\rm ISdust} &=& 
\frac{4\pi}{3} (\vsn t)^{3} n_{\rm ISM} m_{\rm p} f^{-1} \nonumber \\
&\simeq& 2 \times 10^{-6}
\left( \frac{\vsn}{5000\ {\rm km~ s^{-1}}} \right)^3
\left( \frac{t}{30 \ {\rm yr}} \right)^3 \nonumber \\
&& \times \left( \frac{n_{\rm ISM}}{1\  {\rm cm^{-3}}} \right)
\left( \frac{f}{200} \right)^{-1} \Msun.
\label{eq:ISdust}
\end{eqnarray}
The mass of swept-up IS dust is 
nearly $0.1 \Msun$ in SNRs at $t=1000$ yr,
which makes it difficult to estimate the SN dust mass in SNRs.
On the contrary,  the swept-up IS dust mass 
at the transitional phase is only an order of $10^{-6} \Msun$.

In summary, future observations of SNe at the transitional phase will 
the gap of infrared observations of SNe with the age of $10-100$ years.
At $<30$ \mum, observations will provide an interesting 
way to study the circumstellar and interstellar environments of the progenitor.
With SPICA, these components can also be studied by 
low resolution spectroscopy.
If these two components are relatively weak, 
the SN dust might be detected at longer wavelengths, which 
enables the reliable measurement of the dust mass.

\section{Conclusions}
\label{sec:conclusions}

SN explosion have been studied mostly by
observations of extragalactic SNe at
the early phase ($t<$ a few years) or
observations of SNRs at the late phase ($t > 100$ yr).
Observations at the transitional phase from SN to SNR
have not been extensively performed, especially at IR wavelengths.
We present theoretical predictions for the IR emission
from SNe at the transitional phase.
We show that the emission arises from SN dust, 
light echo by CS and IS dust, and shocked CS dust.

We search for IR emission toward 6 core-collapse SNe at the transitional phase
in the nearby galaxies NGC 1313, NGC 6946, and M101
by using the data taken with the AKARI satellite 
and {\it Spitzer}.
Among the targets, we detect emission associated
with SN 1978K in NGC 1313.
The emission is explained by 
$1.3 \times 10^{-3} \Msun$ of silicate dust.
The shocked circumstellar dust 
for gas mass-loss rate $\Mdot = 10^{-4}\ \Msun\ {\rm yr^{-1}}$ 
is the most probable origin of the IR emission from SN 1978K.
Dust formed in the outer, cool dense shell might 
also be a possible origin.

IR emission from the other 5 objects is not detected. 
Current observations are sensitive only to
the total luminosity of $L > 10^{38}\ {\rm erg\ s^{-1}}$.
The non-detection of SN 1962M
excludes the existence of the shocked CS dust 
for the high gas mass-loss rate $\Mdot = 10^{-4}\ {\rm \Msun\ yr^{-1}}$.

Future observations will fill the gap of the IR 
observations at the transitional phase.
At $< 30$ \mum, the emission is
dominated by the shocked CS dust and IS dust echo.
The observations will provide an interesting opportunity 
to study the circumstellar and interstellar environments of the SN progenitor.
SN dust component has a smaller contribution, but 
has a spectrum peaking at longer wavelengths.
If the emission by the shocked CS dust and IS dust echo
does not dominate the IR emission, 
the SN dust might be detected at $> 30$ \mum.
This will give a reliable estimate of the SN dust mass,
which does not suffer from the high optical depth of dust and
the confusion with the interstellar dust.

\acknowledgments
We are grateful to the referee for giving useful comments.
We thank T. T. Takeuchi for providing the model of 
the confusion limit of SPICA.
We also thank T. Kozasa for useful comments, 
and G. Folatelli for valuable discussion.
This research is based on observations with AKARI, 
a JAXA project with the participation of ESA,
and on observations made with the Spitzer Space Telescope, 
which is operated by the Jet Propulsion Laboratory, 
California Institute of Technology under a contract with NASA.
We have made use of data products from the Two Micron All Sky Survey, 
which is a joint project of the University of Massachusetts and 
the Infrared Processing and Analysis Center/California Institute of Technology, 
funded by the National Aeronautics and Space Administration 
and the National Science Foundation.
This research has been supported in part by World Premier
International Research Center Initiative, MEXT,
Japan, and by the Grant-in-Aid for Scientific Research of the 
Japan Society for the Promotion of Science (20340038,22684004,22840009).

\appendix

\section{SEDs for Amorphous Carbon, Graphite, and Forsterite}
\label{app:Mdust_all}

\begin{figure*}
\begin{center}
\includegraphics[scale=1.3]{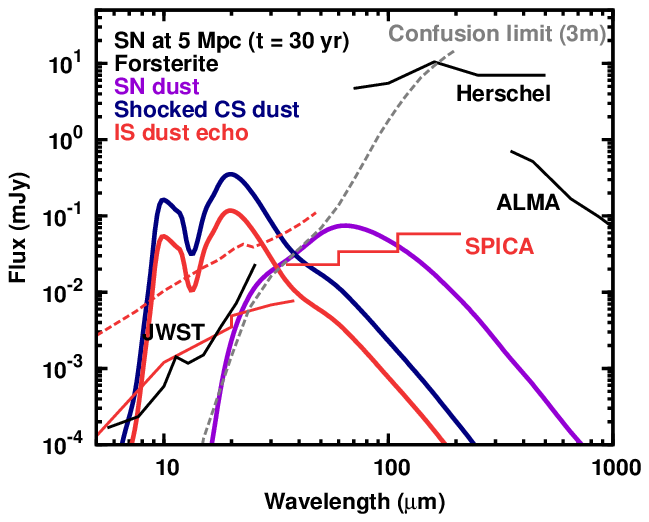}\\
\includegraphics[scale=1.3]{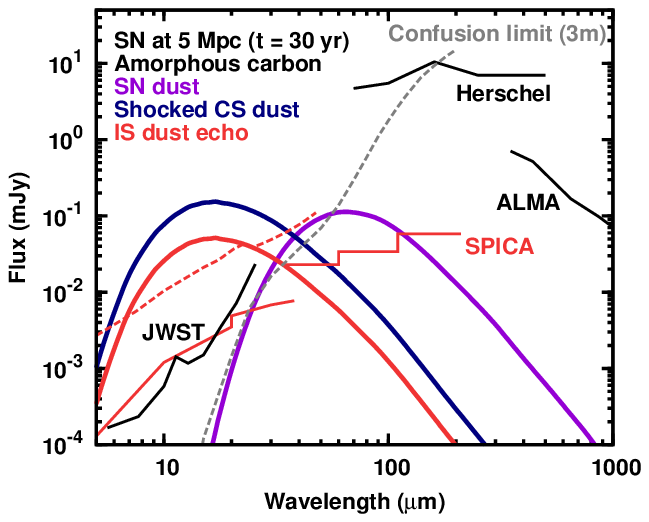}\\
\includegraphics[scale=1.3]{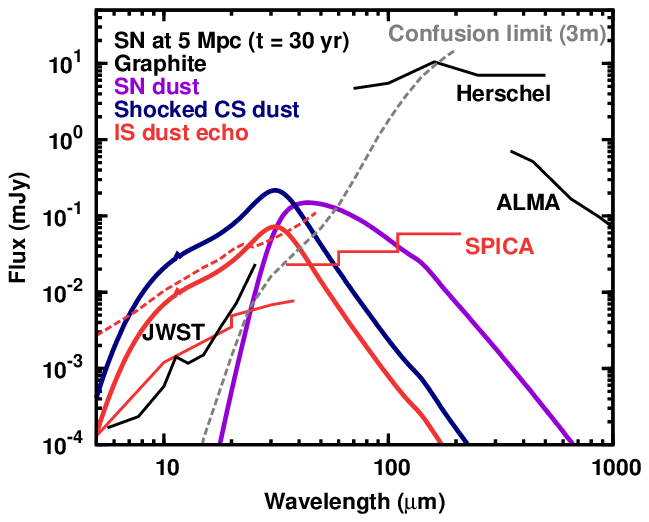}
\caption{
Expected SEDs of SNe at the transitional phase ($t=30$ yr)
for forsterite, amorphous carbon, and graphite (from top to bottom).
For the shocked CS dust, $\Td=100-150$ K and $\Md = 3 \times 10^{-4} \Msun$ 
are assumed.
For the IS dust echo, $\Td = 150$ K and $\Md = 1 \times 10^{-4} \Msun$ 
are assumed, which give the luminosity of $L = 10^{37}\ {\rm erg\ s^{-1}}$.
For the SN dust, $\Td = 50$ K and $\Md = 0.1 \Msun$ are assumed. 
\label{fig:SPICA_all}}
\end{center}
\end{figure*}

As discussed in Section \ref{sec:future},
future mid/far-IR observations will be able to detect the 
shocked CS dust, IS dust echo, and possibly SN dust.
Figure \ref{fig:SPICA} only shows the case for
astronomical silicate.
In Figure \ref{fig:SPICA_all}, we show similar plots,
but for forsterite, amorphous carbon, and graphite.
For each component, the same temperature and dust mass 
with those for astronomical silicate are assumed.

\end{document}